\documentstyle[prl,twocolumn,aps,graphicx]{revtex}

\begin{document}

\twocolumn[\hsize\textwidth\columnwidth\hsize\csname    
@twocolumnfalse\endcsname                               

\begin{title} {\Large \bf
Dipolar Interactions and Origin of Spin Ice in Ising Pyrochlore Magnets}
\end{title} 


\author{Byron C. den Hertog and Michel J. P. Gingras$^{\dag}$}
\address{Department of Physics, University of Waterloo, Ontario Canada
N2L 3G1\\
$^{\dag}$Canadian Institute for Advanced Research}
\date{\today}
\maketitle
\begin{abstract}
Recent experiments  suggest that the Ising pyrochlore magnets
${\rm Ho_{2}Ti_{2}O_{7}}$ and ${\rm Dy_{2}Ti_{2}O_{7}}$ display
qualitative properties of the spin ice model proposed by Harris {\it et al.}
\prl {\bf 79}, 2554 (1997).
We discuss the dipolar energy scale present in both these materials and
consider how  they can display spin ice behavior {\it
despite} the presence of long range interactions.
Specifically, we present  numerical simulations and a mean field analysis of
pyrochlore Ising systems in the presence of nearest neighbor exchange and
long range dipolar interactions.  We find that two possible phases can occur,
a long range ordered antiferromagnetic one  and the other dominated by
spin ice  features. Our  quantitative theory is in very good agreement with
experimental data on both ${\rm Ho_{2}Ti_{2}O_{7}}$ and
${\rm Dy_{2}Ti_{2}O_{7}}$. We suggest that the nearest neighbor
exchange in ${\rm Dy_{2}Ti_{2}O_{7}}$ is {\it antiferromagnetic} and that
spin ice behavior is induced by long range dipolar interactions.

\vspace{-5mm}

\end{abstract}

\vskip2pc]                                              


\narrowtext

\begin{figure}
\begin{center}
\rotatebox{90}{\includegraphics[width=6cm]{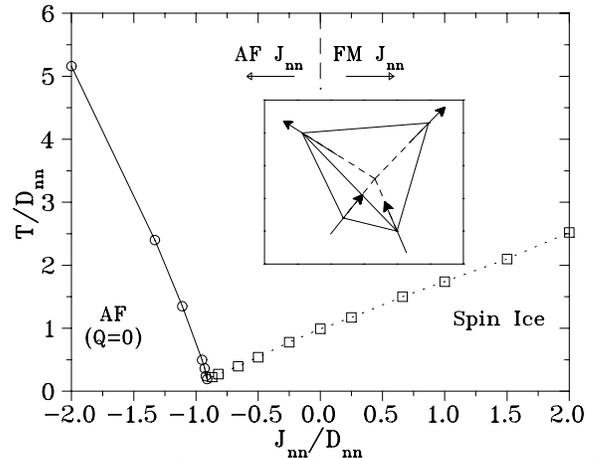}}
\caption{Phase diagram of Ising pyrochlore magnets with nearest
neighbor exchange and long range dipolar interactions.
$D_{{\rm nn}}$ and $J_{{\rm nn}}$ are  the Ising
parameters for the  nearest neighbor dipolar and exchange energies
respectively (see text). {\it Inset:} An ice rule configuration
of a tetrahedron. The four local $\langle 111\rangle$ Ising axes meet in the 
middle of the
tetrahedron [4].}
 \label{phase}
\end{center}
 \end{figure}

An exciting development  has occurred in the last two years  with the 
discovery of an apparent analogy between the low temperature physics of 
the geometrically frustrated Ising pyrochlore compounds 
${\rm Ho_{2}Ti_{2}O_{7}}$ \cite{Harris1} 
and ${\rm Dy_{2}Ti_{2}O_{7}}$  \cite{Ramirez} (so called `spin ice' 
materials), and proton ordering in real ice\cite{Pauling}.
The magnetic cations ${\rm Ho}^{3+}$ and ${\rm Dy}^{3+}$ of these particular
materials reside  on the  pyrochlore lattice of corner sharing tetrahedra. 
Single-ion effects conspire to make their  magnetic moments  almost ideally 
Ising-like, but with their own  set of local axes.  In particular, each moment has its  
local Ising axis along the line connecting  its site to the middle of a tetrahedron to 
which it belongs (see inset of Fig. \ref{phase}).
                               
In  a simple model of nearest neighbor  ferromagnetic (FM) interactions, 
such a  system has the same `ice rules' for the  construction of its ground 
state as those for the  ground state  of real  ice\cite{Pauling,rules}. In  both cases, these 
rules predict a macroscopically  degenerate ground state, a feature that a 
number of  geometrically frustrated systems possess\cite{Villain,Moessner,Gardner,Gingras}.

In ${\rm Ho_{2}Ti_{2}O_{7}}$, $\mu$SR data indicates a lack of ordering down to
$\sim 50$ mK despite a Curie-Weiss temperature $\theta_{{\rm cw}} \sim 1.9$ K, 
while single crystal  neutron scattering data  suggests  the 
development of short-range 
FM correlations, but the  absence of  ordering down to at least 0.35 
K\cite{Harris1}.  ${\rm Ho_{2}Ti_{2}O_{7}}$ also displays field dependent 
behavior consistent
with a  spin ice picture\cite{Harris2}. Quite dramatically, thermodynamic
measurements on ${\rm Dy_{2}Ti_{2}O_{7}}$\cite{Ramirez} show a  lack of 
any ordering feature in the specific heat  data, with the measured
ground state
entropy within 5\% of Pauling's prediction for the entropy of 
ice\cite{Pauling}.

However,  both   spin ice materials  contain  further interactions additional
to the nearest neighbor exchange. Often, rare earth cations can  have 
appreciable magnetic moments and, consequently, magnetic dipole-dipole 
interactions of 
the same order as, if not larger than the exchange coupling,  can occur.  
Furthermore,  it has been suggested  that the nearest neighbor  exchange
interaction  
in ${\rm Ho_{2}Ti_{2}O_{7}}$ is actually {\it antiferromagnetic} (AF) 
\cite{Siddharthan}, which by itself  should cause  a phase transition 
to a long range  
ordered ground state. Thus, {\it how} these systems actually 
display spin ice-like behavior is most puzzling. For example, one might naively expect that the 
long-range and anisotropic spin-space nature of the dipolar 
interactions would introduce so
many constraints that a large degree of the degeneracy present in the simple
nearest-neighbor ferromagnetic spin-ice model\cite{Harris1,rules} 
would be  
removed, and induce long-range order. It is this issue that we wish to address 
in this paper.

A previous attempt to consider dipolar effects in Ising pyrochlores was made
by Siddharthan {\it et al.}\cite{Siddharthan}. In that work, the dipole-dipole 
interaction was truncated  beyond  five  nearest neighbor distances, and
a sharp transition between paramagnetism and a partially ordered phase 
(where rapid freezing occurs)   was observed for interaction 
parameters believed appropriate for 
${\rm Ho_{2}Ti_{2}O_{7}}$\cite{Siddharthan}. However, we argue that 
the truncation of dipole-dipole interactions can be  misleading, 
and  may introduce spurious features in various  thermodynamic properties.
For example, we find that the sharp feature observed in the specific heat   
for truncation  beyond five nearest neighbor distances\cite{Siddharthan} 
is softened and rounded for truncation beyond the tenth nearest neighbor  shell, and the 
observed dynamical freezing  is pushed to lower temperatures.  As  we show 
below, in the limit of infinite range dipoles, the
interaction parameters of Siddharthan {\it et al.} \cite{Siddharthan} yield spin ice.

In this work, we consider the interplay between nearest neighbor exchange and 
dipolar interactions by taking into account the  long range 
(out to infinity) nature of the dipolar interactions through the use of  Ewald 
summation techniques\cite{Born,deLeeuw}.
Our Monte Carlo simulations and mean field results  show that dipolar forces 
are remarkably adept  at producing spin ice physics over a large region of 
parameter space.

Some of our main conclusions are shown in Fig. \ref{phase}.
For Ising pyrochlores, the dipole-dipole interaction at nearest neighbor is 
FM, and therefore favors frustration. Beyond nearest neighbor, the 
dipole-dipole interactions can be either FM or AF, 
or interestingly, even  multiply valued\cite{note}, depending on the neighbor 
distance.  Defining the nearest neighbor dipole-dipole interaction  as 
$D_{{\rm nn}}$ and  the nearest neighbor exchange as $J_{{\rm nn}}$, 
our Monte Carlo results  indicate  that spin ice behavior persists
in the presence of AF exchange up to $J_{{\rm nn}}/D_{{\rm nn}}\sim -0.91$. 
For $J_{{\rm nn}}/D_{{\rm nn}}< -0.91$
we find a second order phase transition to the globally doubly degenerate 
${\bf Q}=0$  phase of the  nearest neighbor AF exchange-only 
model\cite{Bramwell},
where all spins either point  in, or all out of a given tetrahedron.

Our Hamiltonian describing the Ising pyrochlore magnets is as follows,
\begin{eqnarray}
\label{Hamiltonian}
H&=&-J\sum_{\langle ij\rangle}{\bf S}_{i}^{z_{i}}\cdot{\bf S}_{j}^{z_{j}} 
\nonumber \\
&+& Dr_{{\rm nn}}^{3}\sum_{\langle ij\rangle}
\frac{{\bf S}_{i}^{z_{i}}\cdot{\bf S}_{j}^{z_{j}}}{|{\bf r}_{ij}|^{3}} - 
\frac{3({\bf S}_{i}^{z_{i}}\cdot{\bf r}_{ij})
({\bf S}_{j}^{z_{j}}\cdot{\bf r}_{ij})}{|{\bf r}_{ij}|^{5}} \; ,
\end{eqnarray}
where the spin vector ${\bf S}_{i}^{z_{i}}$ labels the Ising moment of 
magnitude $|S|=1$ at lattice site $i$ and {\it local}  Ising axis $z_{i}$. 
Because the local Ising axes belong to the 
set of $\langle 111\rangle$ vectors, the nearest neighbor exchange energy 
between two spins $i$
and $j$ is $J_{{\rm nn}}\equiv J/3$. The dipole-dipole interaction at 
nearest neighbor is $D_{{\rm nn}}\equiv 5D/3$
where $D$ is the usual estimate of the dipole energy scale, 
$D=(\mu_{0}/4\pi)g^{2}\mu^{2}
/r_{{\rm nn}}^{3}$. For both ${\rm Ho_{2}Ti_{2}O_{7}}$ and ${\rm Dy_{2}Ti_{2}O_{7}}$,
$D_{{\rm nn}}\sim 2.35$ K.

It is well known in the field of electrostatic interactions that the 
dipole-dipole interaction 
is difficult to handle due to its $1/r^{3}$ nature. In general,  a lattice 
summation  of dipole-dipole interactions is conditionally convergent, and must
be considered with care. In order to include the important 
long range nature of the dipole-dipole interaction, we have implemented the 
well known Ewald method within our simulation technique, in order to derive an 
{\it effective}\cite{ewaldnote} dipole-dipole interaction between spins within our simulation 
cell\cite{deLeeuw}. Unlike in dipolar fluid simulations, our
lattice spins allow the effective interactions between all moments to be
calculated only once, after which a  numerical simulation can proceed as 
normal.  

A standard Metropolis algorithm was used in our Monte Carlo simulations. We
used a conventional cubic cell for the pyrochlore lattice, which contains
16 spins. In general, we found it sufficient to simulate up to 
$4\times 4\times 4$ 
cubic cells ({\it ie.} L=4, and 1024 spins) with up to $\sim 10^{6}$ Monte 
Carlo steps per spin when necessary. Thermodynamic data were collected by 
starting the simulations at high temperatures and cooling very slowly.

Referring to Fig. \ref{phase}, the characterization of a system as having spin
ice behavior was carried out  by determining the entropy, via numerical
integration of  the  specific heat divided by temperature data. Pauling's
argument for the entropy of ice yields $R[\ln 2 - (1/2)\ln (3/2)]$ or
$4.07$ J mol$^{-1}$ K$^{-1}$\cite{Ramirez,Pauling}. 
We find for $J_{{\rm nn}}/D_{{\rm nn}}=-0.91$, 
(our spin ice data point closest   to the phase
boundary in Fig. \ref{phase}), this value for the entropy to within 
$3\%$, using a  system size $L=4$. 

Our thermodynamic data indicates that when the nearest neighbor exchange is 
AF and large compared to  the dipolar interactions, the system 
undergoes a  second order phase transition  to an all in or all out ${\bf Q}=0$ 
ground state as alluded to 
earlier.  This AF phase persists slightly beyond the point where the
nearest neighbor dipolar interaction (FM) is stronger than nearest 
neighbor AF exchange.

\begin{figure}
\begin{center}
\rotatebox{90}{\includegraphics[width=5.5cm]{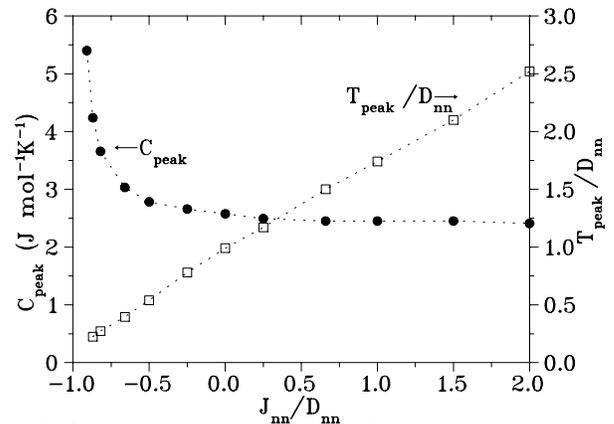}}
 \caption{Dependence of  the specific heat peak height and temperature 
location on exchange and dipole-dipole interaction parameters.}
 \label{peak}
\end{center}
 \end{figure}

In the spin ice regime, our specific heat data has  a number of interesting 
features which help shed light on the effect of long range dipole-dipole 
interactions. In general, each data set  shows qualitatively the same 
broad peak as observed in the  nearest neighbor FM exchange model, 
and vanishes at high and low temperatures\cite{Harris2,Siddharthan}. We
detect very little system size dependence upon comparison of data from
 $L=2,3,4$ 
simulation sizes. In particular, there is no noticeable size dependence of
 the specific heat maximum, nor its position.
As the AF/spin ice phase boundary is approached from the spin ice 
side, the specific heat peak begins to narrow and more importantly, both the 
peak height 
and peak position begin to vary. As we discuss below, this has important 
ramifications for the interpretation of experimental data.

In Fig. \ref{peak} we plot the dependence of the specific heat peak height 
($C_{{\rm peak}}$) and the temperature  at which it occurs ($T_{{\rm peak}}$) 
on the ratio
of the nearest neighbor exchange and dipole-dipole  energies, 
$J_{{\rm nn}}/D_{{\rm nn}}$. 
Note that in the regime of large nearest neighbor FM coupling, 
the peak height plateaus to the value one observes in the nearest neighbor 
FM exchange-only model.

\begin{figure}
\begin{center}
\rotatebox{90}{\includegraphics[width=6cm]{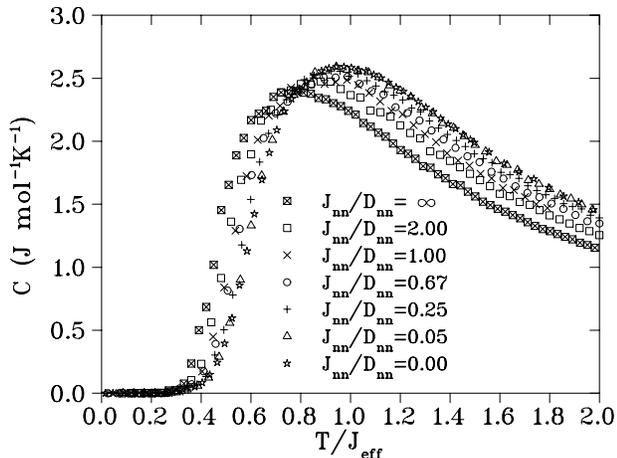}}
 \caption{Specific heat for system size $L=2$ rescaled into units of the 
effective nearest neighbor interaction $J_{{\rm eff}}$. 
$J_{{\rm nn}}/D_{{\rm nn}}=0$ corresponds to purely dipolar interactions while
$J_{{\rm nn}}/D_{{\rm nn}}=\infty$ corresponds to nearest neighbor FM 
exchange only.} 
 \label{scale}
\end{center}
 \end{figure}

 Indeed our data suggests that when the exchange becomes FM,
the nearest neighbor effective bond energy is large enough to dominate the excitations
of the system. This can be more dramatically seen in Fig. \ref{scale}, where
we have rescaled the specific heat curves for a number of interaction parameter
values in terms of the effective nearest neighbor interaction 
$J_{\rm{eff}}\equiv J_{{\rm nn}}+D_{{\rm nn}}$ in the
regime $J_{{\rm nn}}/D_{{\rm nn}}>0$. 

This figure shows that in terms of an effective energy scale, the medium to
long range effects of the dipolar interactions are  `screened' by
the system, and one recovers qualitatively the short range physics of the
nearest neighbor spin ice model. 
Remarkably, inclusion of long range dipolar interactions appears to 
have the effect of {\it removing} a tendency towards ordering which can come 
from a short range truncation of the dipole-dipole interaction.

As the nearest neighbor exchange interaction becomes AF, we find that
the approximate `collapse' onto a single energy scale becomes less accurate, with
features in the specific heat  becoming dependent on $J_{{\rm nn}}/D_{{\rm nn}}$ in a 
more complicated manner. It is within this regime that we believe that both 
${\rm Ho_{2}Ti_{2}O_{7}}$ and ${\rm Dy_{2}Ti_{2}O_{7}}$ exist.

Since $D_{{\rm nn}}$ is a quantity which can be calculated once the crystal field 
structure of the magnetic ion is known, the nearest neighbor exchange 
$J_{{\rm nn}}$ is  the only 
adjustable parameter in our theory. Fig. \ref{peak} enables us to test in two 
independent ways the usefulness of our approach to the long range dipole 
problem in spin ice materials. 

\begin{figure}
\begin{center}
\includegraphics[width=8cm]{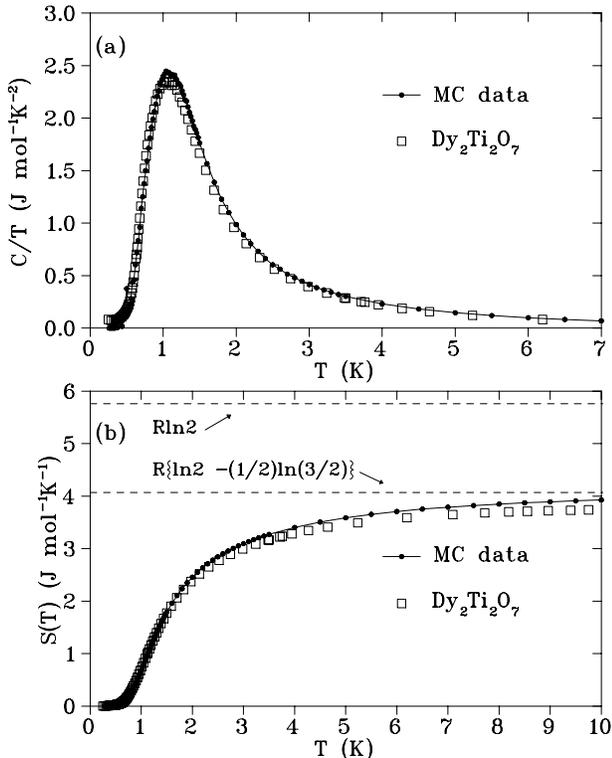}
\caption{Comparison of  (a) specific heat  and (b) entropy data between 
 ${\rm Dy_{2}Ti_{2}O_{7}}$ [2]  and  Monte Carlo simulation  
with $J_{{\rm nn}}=-1.24$ K,  $D_{{\rm nn}}=2.35$ K and  system size $L$=4. } 
 \label{dyent}
\end{center}
 \end{figure}

If we consider ${\rm Dy_{2}Ti_{2}O_{7}}$ for
example, specific heat measurements by Ramirez {\it et al.} \cite{Ramirez} 
indicate a  peak height, $C_{\rm peak}^{\rm Dy}$ of
\mbox{$\sim 2.72$ J mol$^{-1}$K$^{-1}$}. Given that 
\mbox{$D_{{\rm nn}}\approx 2.35$ K} for 
this material, the left  hand plot of \mbox{Fig. \ref{peak}} indicates a nearest neighbor 
exchange coupling \mbox{$J_{{\rm nn}}\sim -1.2$ K}. The same  experimental 
specific heat data shows that this   peak occurs at a temperature of  
$T_{{\rm peak}^{\rm Dy}} \sim 1.25$ K. 
Using the   plot of $T_{{\rm peak}}$ in Fig. \ref{peak}, we independently arrive at 
approximately the same  conclusion for the  value of the nearest neighbor 
exchange. Thus, we predict that AF  exchange is present in 
${\rm Dy_{2}Ti_{2}O_{7}}$ with $J_{{\rm nn}}\sim -1.2$ K.  If there was no
AF exchange present in this system, our results in Fig. \ref{phase} imply that
there would be a peak in the specific heat at a  temperature of at least
$\sim 2.3$ K, which is not observed experimentally\cite{Ramirez}.
Our best fit for the specific heat data of ${\rm Dy_{2}Ti_{2}O_{7}}$ by     
Ramirez {\it et al.}\cite{Ramirez}   is shown in Fig. \ref{dyent}, 
where we find good agreement between theory and experiment for 
$J_{{\rm nn}}=-1.24$ K.
\vspace{-4mm}
Specific heat measurements on a 
powdered sample of ${\rm Dy_{2}Ti_{2}O_{7}}$ in a magnetic field were also
reported in Ref. \cite{Ramirez}. Three field independent peaks were observed 
at low temperature. For a large field in the $\langle 110\rangle$ direction,
two  spins on each tetrahedron are pinned by the field, while the other
two remain free, since their Ising axes are perpendicular to the applied field.
 Due to the dipolar interaction, there will be a
coupling between  the  fluctuating spins on these two sub-lattices. Our 
preliminary simulations on small lattice sizes suggest a field independent 
ordering at low temperature as observed in experiment.

Considering ${\rm Ho_{2}Ti_{2}O_{7}}$, experimental data on its thermodynamic
properties is not so categorical. The specific heat data of Siddharthan 
{\it et al.} \cite{Siddharthan}  indicates  a feature at $\sim 0.8$ K, 
although it has been suggested that this could be due to  an additional  
contribution in 
this temperature range from an anomalously large hyperfine coupling in 
${\rm Ho}^{3+}$\cite{Bramwell2}. Nevertheless, using the plot of 
$T_{{\rm peak}}$ in
Fig. \ref{peak}, we find substantial AF exchange coupling of the same order
of magnitude as in the Dy compound. We note that  Siddharthan 
{\it et al.}\cite{Siddharthan}  and den Hertog {\it et al.}\cite{Bramwell2} 
find from 
analysis of magnetization measurements a similar order of magnitude for 
$J_{{\rm nn}}$. 
Furthermore, our numerical simulations within this region of parameter space
indicate a Curie-Weiss temperature of $\sim 2$ K, in agreement with an
experimental estimate by Harris {\it et al.} of $\theta_{{\rm cw}}\sim 1.9$ K\cite{Harris1}.

 ${\rm Tb_{2}Ti_{2}O_{7}}$ is an Ising pyrochlore system of similar type to
the Ho$^{3+}$ and Dy$^{3+}$ based materials\cite{Gardner,Gingras}, but the 
Ising anisotropy is reduced to much lower temperature due to narrowly spaced
 crystal field levels\cite{Gingras}. 
While this makes the interpretation of experimental data more 
difficult, initial estimates of the nearest neighbor exchange and dipole
moment\cite{Gingras} yield $J_{{\rm nn}}/D_{{\rm nn}}\sim -1$, placing this system very 
close to the phase boundary of \mbox{Fig. \ref{phase}}. 
Indeed, $\mu$SR measurements suggest that 
 ${\rm Tb_{2}Ti_{2}O_{7}}$ fails to order down to 70 mK\cite{Gardner} and thus 
it would appear   that
a spin ice picture for this material cannot be  {\it a priori} ruled out.
 
While we believe our approach yields a reasonably successful quantitative 
theory of spin ice behavior in Ising pyrochlores, there still remains the
question of why long range dipolar interactions do not appear to  lift the
macroscopic degeneracy associated with  the ice rules, and select an 
ordered state.

Mean-field theory  provides a more quantitative basis for examining this issue. Following the
approach used in Refs.\cite{Reimers,Raju}, we find that the soft-modes for
Ising pyrochlore systems described by Eq. \ref{Hamiltonian} consist of two 
very weakly dispersive branches (less
than 1\% dispersion) over the whole Brillouin zone (except at ${\bf
Q}\equiv 0$). Such  a set of quasi-dispersionless branches is very similar to 
the two completely dispersionless soft branches of the nearest-neighbor
FM spin ice model. Consequently, both nearest-neighbor
FM and dipolar spin ice behave almost identically over the
whole temperature range spanning $O(w) \leq T < \infty$ where $w$ is the
bandwidth of the soft branch of dipolar spin ice. We note that this
near dispersionless behavior is only recovered {\it asymptotically} as 
the long range dipoles are included out to infinity.       
The lifting of degeneracy at the 1\% level in the apparent
absence of any small parameters in
the theory is at this point not completely understood. We do not
believe it is due to numerical or computational error.
A partial explanation may be that $(111)$ Ising anisotropy,
the $1/r^3$ long-range nature of dipolar interactions,
and the  specific relationship between the topology of the pyrochlore lattice
and the anisotropic (spin$-$space) coupling of dipolar interactions, 
combine in a subtle manner to produce a spectrum of soft modes that
{\it approximate} very closely
the spectrum of the nearest-neighbor
ferromagnetic spin ice model of Harris et al.\cite{Harris1}. 
In other words, there
must be an almost exact symmetry fullfilled in this system when 
long-range dipolar interactions are taken into account.
However, the same long-range nature of these interactions renders it
difficult to construct a simple and intuitive picture of their effects,
and we have not been able to identify such ``almost exact'' symmetry.

Also, in   these systems,   the absence of
soft fluctuations (Ising spins) combined with the macroscopic degeneracy  
associated with  the  ice rules  may be such  that
 correlations associated with a  `true' ground state are
dynamically inhibited  from developing. For energetic  reasons,  
states obeying the ice rules are
favored down to low temperature, by which time such large energy barriers 
exist that evolution towards the true ground state  is never achieved. In 
simulation terms, at low temperature the Boltzmann weights for local spin 
flips `towards' this state are simply too small.

In conclusion,  using  Ewald summation techniques we have 
considered the effects of long range dipole-dipole interactions
on the magnetic behavior of Ising pyrochlore systems. Our  
results show that spin ice behavior is recovered over a large region of 
parameter space, and we find  quantitative agreement between our approach and
experimental data  on spin ice materials. 

We  thank   S. Bramwell, B. Canals, and P. Holdsworth
for  useful discussions. We are grateful to  A. Ramirez for making
available his specific heat data. M.G. acknowledges financial support from 
NSERC of Canada, Research Corporation for a Research Innovation Award and a 
Cottrell Scholar Award, and the Province of Ontario for a Premier Research 
Excellence Award.


\begin{references}
\bibitem{rules}
For the particular type of Ising pyrochlores discussed here, the ice rules
imply that every tetrahedron must have two spins pointing into it, and two
pointing out.
\bibitem{ewaldnote}
 Physically, this corresponds to having an infinite number
of replicas of the simulation cell, and using the Ewald  method to
sum convergently the dipolar interaction between say  spins $i$ and $j$ 
(within the simulation cell) and spin $j$'s images 
(outside the simulation cell). The result, is the effective Ising interaction 
of spin $i$ with spin $j$.
\end{references}
\end{document}